\documentclass[sn-mathphys]{sn-jnl}

\usepackage{graphicx}
\usepackage{dcolumn}
\usepackage{bm}
\usepackage{comment}
\usepackage{mathtools} 
\usepackage{amsmath}
\usepackage{amssymb}
\usepackage{tabularx}
\usepackage[utf8]{inputenc}
\usepackage{ifthen}
\usepackage{breqn}
\jyear{2021}%

\theoremstyle{thmstyleone}%
\theoremstyle{thmstyletwo}%

\theoremstyle{thmstylethree}%
\usepackage{mathtools}

\begin{document}

\title[Ab initio Approaches to High Entropy Alloys]{Ab initio Approaches to High Entropy Alloys:}
\subtitle{A Comparison of CPA, SQS, and Supercell Methods}


\author*[1]{\fnm{Mariia} \sur{Karabin}}\email{karabinm@ornl.gov}

\author[2]{\fnm{Wasim} \sur{Mondal}}

\author[6]{\fnm{Andreas} \sur{\"Ostlin}}

\author[5]{\fnm{Wai-Ga D.} \sur{Ho}}

\author[5]{\fnm{Vladimir} \sur{Dobrosavljevic}}

\author[3]{\fnm{Ka-Ming} \sur{Tam}}

\author[2]{\fnm{Hanna} \sur{Terletska}}

\author[6]{\fnm{Liviu} \sur{Chioncel}}

\author[4]{\fnm{Yang} \sur{Wang}}
\author*[1]{\fnm{Markus} \sur{Eisenbach}}\email{eisenbachm@ornl.gov}


\affil*[1]{\orgdiv{Advanced Computing for Chemistry and Materials}, \orgname{Oak Ridge National Laboratory}, \orgaddress{\city{Oak Ridge}, \state{TN} \postcode{37831}, \country{USA}}}

\affil[2]{\orgdiv{Department of Physics and Astronomy}, \orgname{Middle Tennessee State University}, \orgaddress{\city{Murfreesboro}, \state{TN} \postcode{37132} \country{USA}}}

\affil[3]{\orgdiv{Department of Physics and Astronomy}, \orgname{Louisiana State University}, \orgaddress{\city{Baton Rouge}, \state{LA} \postcode{70803}, \country{USA}}}

\affil[4]{\orgdiv{Pittsburgh Supercomputing Center}, \orgname{Carnegie Mellon University}, \orgaddress{\city{Pittsburgh}, \state{PA} \postcode{15213}, \country{USA}}}

\affil[5]{\orgdiv{Department of Physics}, \orgname{Florida State University}, \orgaddress{\city{Tallahassee}, \state{FL} \postcode{37132}, \country{USA}}}

\affil[6]{\orgdiv{Center for Electronic Correlations and Magnetism, Institute for Physics}, \orgname{University of Augsburg}, \orgaddress{\city{Augsburg}, \country{Germany}}}






\abstract{
We present a comparative study of different modeling approaches to the  electronic properties of the $\textrm{Hf}_{0.05}\textrm{Nb}_{0.05}\textrm{Ta}_{0.8}\textrm{Ti}_{0.05}\textrm{Zr}_{0.05}$ high entropy alloy. Common to our modeling is the methodology to compute the one-particle Green's function in the framework of density functional theory. We demonstrate that the special quasi-random structures modeling and the supercell, i.e. the locally self-consistent multiple-scatering methods provide very similar results for the ground state properties such as the spectral function (density of states) and the equilibrium lattice parameter. To reconcile the multiple-scattering single-site coherent potential approximation with the real space supercell methods, we included the effect of screening of the net charges of the alloy components.   
Based on the analysis of the total energy and spectral functions computed within the density functional theory, we found no signature for the long-range or local magnetic moments formation in the $\textrm{Hf}_{0.05}\textrm{Nb}_{0.05}\textrm{Ta}_{0.8}\textrm{Ti}_{0.05}\textrm{Zr}_{0.05}$ high entropy alloy, instead we find possible superconductivity below $\sim 9$K.

}\footnote{{\footnotesize \noindent This manuscript has been authored in part by UT-Battelle, LLC, under contract DE-AC05-00OR22725 with the US Department of Energy (DOE). The US government retains and the publisher, by accepting the article for publication, acknowledges that the US government retains a nonexclusive, paid-up, irrevocable, worldwide license to publish or reproduce the published form of this manuscript, or allow others to do so, for US government purposes. DOE will provide public access to these results of federally sponsored research in accordance with the DOE Public Access Plan (\url{http://energy.gov/downloads/doe-public-access-plan}).}}




\maketitle

\section{Introduction}\label{sec1}


Order or rather disorder, lies at the heart of the development of advanced materials. The traditional design of novel materials starts from a primary or host material, to which other components are added to adjust the performance of the material. Many of modern alloys with attractive properties were found to be located in the centers of the phase diagrams rather than close to its corners. These materials are not formed by a single host and a minor component, instead they contain multiple elements and crystallize as solid solutions. 

The structural order/disorder can be characterized through the entropy which is a measure of the randomness in the system, and it turned out to be an important factor in the fabrication of unordered intermetallics such as high-entropy alloys (HEA).  
%
The most common for these alloys are simple structures (like body-centered cubic (bcc)) with extremely high chemical disorder. Typically, HEA are composed of four or more components to achieve high entropy of mixing, which stabilize the crystal structure and results in highly tunable properties~\cite{george2019high,miracle2017critical,tsai2014high,george2020high,cantor2004microstructural}. In this paper our system of choice is the class of HfNbTaTiZr refractory high entropy alloys, as they have a great potential for a variety of applications. This system has been previously synthesized and studied for biocompatibility and wear resistance for applications in the biomedical field\cite{Mot2019}, high temperature mechanical properties for potential applications in aerospace and power-generation industries \cite{Senkov2012}, and superconducting properties\cite{ko.vr.14}.

As the random placement of atomic species on the lattice sites destroys the translational symmetry that underpins most first-principles calculations of solid state systems, multiple methods have been proposed to deal with disordered alloy systems. Here we will provide a comparison of different approaches for the first principles calculations of alloys to allow the informed decision to chose a computational method for these materials. 

Density functional theory (DFT)~\cite{PhysRev.136.B864,PhysRev.140.A1133} methods have been applied in the theoretical studies of HEA systems, see for instance the recent review~\cite{ik.gr.19}.  
Note that the presence of multi-components leads to a ``reduction'' of the translation symmetry nevertheless the underlying crystalline space group symmetry of the lattice still exists. 
This allows in fact a supercell description which constitutes one of the possibility to model a given disorder realization.
Alternative to the supercell methods, which includes both large random cells as well as the special quasi-random structure (SQS) model, is the class of effective medium theories such as the Coherent Potential Approximation (CPA).
In fact, any method using an artificial partitioning of the crystal into regions associated with particular alloy components, may possess nonzero net charges. This is by now a well known problem originating form the conventional effective medium single-site approximation and the self-consistent DFT loop in computing the electrostatic contribution through the Poisson equation. 
The correct solution of the Poisson equation should reflect the charge distribution in the lattice, 
which leads to an apparent inconsistency between the CPA (producing a homogeneous charge distribution) and the electrostatic part of the DFT (favoring inhomogeneity).
A way out of this problem is to modify the electrostatic potential for each alloy component in agreement with the assumptions of the CPA.
In practice this amounts to introduce an additional shift of the one-electron potential inside each atomic sphere representing the interaction of the electrons within the sphere with the missing charge distributed outside the sphere. Such a correction will be applied in the Korringa-Kohn-Rostoker CPA calculations and allows to obtain a very good agreement with the supercell and SQS methods.  

In this paper we present a comparison of the three \textit{ab initio} approaches to disordered metallic systems applied to high entropy alloys: the effective medium CPA method, the supercell method, and the special quasi-random structure (SQS). All three approaches presents advantages and disadvantages~\cite{Tian2017}. The CPA is less computationally expensive, when compared to the SQS and the Supercell methods, since only a small unit cell is required to represent any arbitrary composition of HEA. However, the SQS and Supercell methods capture correctly the local chemical environment effects through the long range electrostatic interactions. 
Our comparative study applied to HfNbTaTiZr extends over the total energy computations, spectral functions and charge distributions. The total and alloy component resolved  density of states show a local maxima around the Fermi level. Thus, we have searched for the existence of possible magnetic instabilities or formation of local moments. Our spin-polarized DFT-LSDA calculations and the DFT based disordered local moment (DLM) applied to HfNbTaTiZr did not show indication for the existence of long-range or local magnetic moments. Instead based on the knowledge of the existence of superconductivity in this family of compounds, we have estimated the possible critical temperature of this alloy based on the Gaspary-Gyorffy formula~\cite{ga.gy.72} to $\approx 9.3$~K considering the similar Debye temperature as the experimentally reported  Hf$_{0.08}$Nb$_{0.33}$Ta$_{0.34}$Ti$_{0.11}$Zr$_{0.14}$ alloy~\cite{ko.vr.14}.



\section{Electronic Structure Using Green's Functions}

In this section, we give a brief description of the \textit{ab initio} Green's function based approaches to random alloys. We start with the general description of the Korringa-Kohn-Rostoker (KKR)-based methods, followed by the description of two classes of the numerical approaches to get the KKR solution. First, we described the effective medium approaches (the KKR Coherent Potential Approximation (KKR-CPA) and its extension), and second, we describe the supercell methods (the locally self-consistent multiple scattering (LSMS) and a Special Quasi-Random structure approach).

\subsection{Multiple Scattering and Korringa-Kohn-Rostoker Methods}

Many computational schemes use single-particle Kohn-Sham (KS) eigenfunctions and KS eigenvalues~\cite{PhysRev.140.A1133} to obtain the density and the ground-state energy, respectively. Within the Korringa–Kohn–Rostoker Green’s function (KKR-GF) method~\cite{Kohn_1954,Korringa_1947} the single-particle Green's function (GF) of the KS equation is constructed
providing the equivalent information. This method is not restricted to periodic solids but can also be applied also to finite or even disordered systems. Specific to this method is that instead of a Hamiltonian $H$, the resolvent of $H$, i.e. the GF plays the central role.
In the KKR method, the system is partitioned into cells, each of which centered around an atom. For a cell $n$ with volume $\Omega_n$ whose center is the  atomic site with the position vector $\bf{R}_n$, the local potential is given as
\begin{equation}\label{loc_pot}
    v_n(\bf{r}_n)=\begin{cases} V_{\rm eff}({\bf r}), \quad {\rm if} \, \bf{r} \in \Omega_n; \\
    0, \quad \, \quad  \quad {\rm otherwise;}
    \end{cases}
\end{equation}
here ${\bf r}_n={\bf r}-{\bf R}_n$, and $V_{\rm eff}$ is a sum of localized one-electron potential, $v_n$, within each cell. Having chosen a decomposition of space (muffin-tin, or Wigner–Seitz construction) a scalar/fully-relativistic single-electron Hamiltonian is constructed including the Hartree potential and the exchange-correlation potential with local (spin) density approximation (LDA)~\cite{PhysRev.140.A1133,Barth1972} or generalized gradient approximation~\cite{PhysRevB.45.13244,PhysRevLett.77.3865}. 
The resulting single-site problem is solved numerically providing energy-dependent scattering solutions for the isolated potentials located at the atomic sites in terms of solution of the 
Lippmann-Schwinger equation with respect to an unperturbed Bloch wave.
For a potential which is zero outside some domain (sometimes spherical for the spherical potential approximation) the radial wave function inside the potential region is computed in terms of spherical Bessel and Hankel functions. 
From the radial solutions the single-site $t$-matrix, representing the scattering behaviour at a single-site potential, is constructed and the boundary conditions are imposed in terms of  the regular and irregular solutions, which allows also to set-up the full GF.

Among the various techniques available to calculate the GF, the Multiple Scattering Theory (MST) is especially appealing because it splits the solution of the problem into a potential (single site)- and geometry-related (multiple sites) form.
The single-site scattering involves the $t$-matrix, while the multiple scattering problem is solved using the scattering path-operator, i.e. the corresponding scattering matrix for a given finite or infinite array of scatterers is given as
\begin{equation}
    \underline{\tau}^{nm}(\epsilon)=\underline{t}^{n}(\epsilon)\delta_{nm}+\underline{t}^n(\epsilon)\sum_{k\neq n}\underline{g}^{nk}(\epsilon)\underline{\tau}^{km}(\epsilon)
\end{equation}
here $\underline{t}^n(\epsilon)$ is a single site scattering $t$-matrix associated with the potential $v_n$, and $\underline{g}^{nk}(\epsilon)$ is the free-electron propagator matrix. The calculation of the Kohn-Sham orbital wave function is unnecessary in the KKR-GF approach, as the spectral function and dispersion can be obtained directly from the GF constructed using the multiple scattering path operator
\begin{equation}
    G({\bf r}_n,{\bf r}_m;\epsilon)=\sum_{L,L'}Z_{L}^{n}({\bf r}_n;\epsilon)\tau_{LL'}^{nm}(\epsilon)Z_{L'}^{m*}({\bf r}_m;\epsilon)-\sum_L Z_L^n ({\bf r}_n;\epsilon) J_L^{n*} ({\bf r}_n;\epsilon)\delta_{nm},
\end{equation}
where $Z_L^n({\bf r}_n;\epsilon)$ and $J_L^n({\bf r}_n;\epsilon)$ represent the regular and irregular local solutions to Schr\"odinger equation for a single potential Eq.~\ref{loc_pot} in cell $n$ and for the angular momentum indices $L=(l,m)$. Once the GF becomes available, the electronic charge density in the atomic cell $\Omega_n$ can be easily calculated by taking the imaginary part of the GF as follows,
\begin{equation}
    \rho^n({\bf r}_n) = -\frac{2}{\pi}\mathfrak{Im} \int_{-\infty}^{\epsilon_{\rm F}}d\epsilon\,G({\bf r}_n,{\bf r}_n;\epsilon),
\end{equation}
where the factor 2 arises from the two spin states, and $\epsilon_{\rm F}$ is the Fermi energy.

\subsection {
Embedding via an Effective Medium Approach}

\subsubsection{KKR-CPA}
The coherent potential approximation (CPA) combined with the KKR method provides a powerful first principles technique for random alloy systems. The KKR-CPA approach is an effective medium approach, which is based on the assumption that the random system may be mapped onto the impurity placed in the ordered effective medium which is determined self-consistently.~\cite{Soven_1967,Shiba_1971,Gyorffy_1972,Stocks_1978} The CPA medium can be imagined as a periodic system consisting of the ``virtual" species, described by effective medium $t-$matrix $\underline{t}_{\rm CPA}(\epsilon)$. The CPA effective medium scattering path matrix $\underline{\tau}_{\rm CPA}$ is then given as 
\begin{equation}
    \underline{\tau}_{\rm CPA}^{nm}(\epsilon)=\underline{t}_{\rm CPA}^{nn}(\epsilon)\delta_{nm}+\underline{t}_{\rm CPA}^{nn}(\epsilon)\sum_{k\neq n} \underline{g}^{nk}(\epsilon)\underline{\tau}^{km}_{\rm CPA}(\epsilon).
\end{equation}
The local site-diagonal part of the effective medium scattering path operator, then can be formally rewritten as 
\begin{equation}
    \underline{\tau}_{\rm CPA}^{nn}(\epsilon)=[\underline{t}^{-1}_{\rm CPA}(\epsilon)-\underline{\Delta}_{\rm CPA}^{nn}(\epsilon)]^{-1},
      \label{tau_cpa_delta}
\end{equation}
where $\underline{\Delta}_{\rm CPA}^{nn}$ is a renormalized interactor, which is independent of the nature of the potential at a site $n$. 

The CPA medium is then determined self-consistency by placing the actual impurity of species $\alpha$ in the medium. For the impurity of species $\alpha$ at site $n$, the impurity multiple scattering path matrix $\underline{\tau}_{\alpha}^{nn}$ is calculated by replacing the effective medium $t-$matrix by a real impurity $t-$matrix, $\underline{t}_{\alpha}$
\begin{eqnarray}
    \underline{\tau}_{\alpha}^{nn}(\epsilon) & =& [\underline{t}^{-1}_{\alpha}(\epsilon)-\underline{\Delta}_{\rm CPA}^{nn}(\epsilon)]^{-1}\nonumber \\
   \underline{\tau}_{\alpha}^{nn}(\epsilon)^{-1} &=& {\underline{\tau}_{\rm CPA}^{nn}(\epsilon)}^{-1}+{\underline{t}_{\alpha}^{n}}(\epsilon)^{-1}-{\underline{t}_{\rm CPA}^{n}}(\epsilon)^{-1},
\label{tau_alpha}
\end{eqnarray}
The KKR-CPA self-consistency condition requires that  
\begin{equation}
    \underline{\tau}_{\rm CPA}^{nn}(\epsilon)=\sum_{\alpha}c_{\alpha}\underline{\tau}_{\alpha}^{nn}(\epsilon)
       \label{tau_imp}
\end{equation}
which imposes the condition that the replacement of an effective $t-$ matrix by the impurity $t_{\alpha}$ matrix should, on average, produce no changes to the medium. 
Furthermore, since the CPA medium is translation-invariant, the $t-$matrix and the multiple-scattering path matrix $\tau$ for the medium are related through the equation 
\begin{equation}
    \underline{\tau}_{\rm CPA}^{nn}(\epsilon)=\frac{1}{\Omega_{BZ}}\int [\underline{t}_{\rm CPA}^{-1}(\epsilon)-\underline{g}({\bf k},\epsilon)]^{-1},
     \label{tau_BZ}
\end{equation}
where $\Omega_{BZ}$ is the first Brillouin zone volume, and $\underline{g}({\bf k},\epsilon)$ is the lattice Fourier transform of the free electron propagator $\underline{g}^{nk}(\epsilon)$. In the CPA self-consistency loop, Equations (\ref{tau_cpa_delta})-(\ref{tau_BZ}) are combined such that $\underline{t}_{\rm CPA}(\epsilon)$ is determined iteratively.~\cite{Rowlands_2009}

\subsubsection{KKR-CPA and the Electrostatic Potential Correction
}\label{CPA-Correction}

As described in the previous section, the conventional KKR-CPA method is a single site local approximation.
The CPA medium is obtained self-consistently by taking the average of a single impurity structures consisting of one chemical species and its surrounding CPA ``atoms," described by single scattering $t$-matrix $\underline{t}_{\rm cpa}$. For a self-consistent field (SCF) calculation in the conventional KKR-CPA method, such single impurity structures are also applied in the determination of the electrostatic potential: each chemical species experiences the same long range electrostatic interaction,  arising from an averaged environment which is considered charge neutral. Especially, when atomic sphere approximation (ASA) is applied, such long range electrostatic interaction is essentially zero. Despite being appealing in physics intuition and straightforward to implement in practice, the ansatz that on average, each chemical species ``feels" the same long-range electrostatic potential turns out to be a questionable approximation. 

As shown by supercell calculations, the long range electrostatic potential acting at each site is found to be almost linearly dependent on the net charge of the atom at the site, known as the qV relation: The potential versus the net charge of the atoms of the same species type falls approximately on a straight line. One important observation is that the long-range electrostatic potential, averaged over the atoms of the same species, turns out to be different for different species, rather than being the same as assumed by conventional KKR-CPA method, which obviously needs to be corrected to incorporate this observation. In addition to the qV relation, several electrostatic potential models have also been proposed for random alloys~\cite{PhysRevB.42.11388,ABRIKOSOV1992867,PhysRevB.48.11553}. Among those potential models, the charge screening (CS) model~\cite{PhysRevB.48.11553,PhysRevB.66.024201} is the most straightforward to implement in the KKR-CPA method. In the CS model, the net charge at an atomic site is assumed to be completely screened in the nearest neighbor shell, so that the electrostatic potential arising from the environment is determined by the total charge on the neighbor shell which is equal in magnitude but opposite in sign to the net charge of the atom at the given site. Therefore, based on the CS model, an extra potential needs to be added to to the effective one-electron potential for species $\alpha$, 
\begin{equation}
\label{eq:CPA_CS}
    \Delta V^{\rm CS}_\alpha = -e^2\frac{\Delta Q_\alpha}{R_1},
\end{equation}
where $\Delta Q_\alpha$ is the average net charge of the species and $R_1$ is the nearest neighbor shell distance of the site. The contribution of this electrostatic potential to the total energy is obviously given by   
\begin{equation}
    \Delta E^{\rm CS} = -\frac{e^2}{2}\sum_\alpha c_\alpha \frac{\Delta Q^2_\alpha}{R_1}.
\end{equation}
An alternative approach to the inclusion of long-range electrostatic interaction in the KKR-CPA method is to apply the qV relation by letting 
\begin{equation}
\label{eq:CPA_qV}
    \Delta V^{\rm qV}_\alpha = A_\alpha \Delta Q_\alpha + B_\alpha,
\end{equation}
where $A_\alpha$ and $B_\alpha$ are obtained from a supercell calculation via, e.g., LSMS method, and the total energy contribution from this potential correction is
\begin{equation}
    \Delta E^{\rm qV} = -\frac{1}{2}\sum_\alpha c_\alpha \Delta V^{\rm qV}_\alpha \Delta Q_\alpha - {\rm d.c.},
\end{equation}
where ``d.c." is the double counting term that needs to be removed from the total energy. Careful analysis shows that this double counting term exists for the calculations with muffin-tin approximation, but is zero for the ASA calculations~\cite{PhysRevB.66.024201}.

\subsection{Supercell Methods
}
\subsubsection{The Locally Self-Consistent Multiple Scattering (LSMS) Method}
To judge the validity of the various approximation approaches for alloys, it is desirable to consider calculations that as best as possible represent the actual arrangement of atomic species in the material. For the substitutional alloys we are investigating here, this can be achieved with large supercell calculations with the sites randomly occupied by the chemical elements. With sufficiently large supercells, this approach can capture the fluctuations in the distribution of the occupancy of neighbour sites. Additionally, cells could be constructed to capture systems with different short range order. A supercell calculation in the KKR method requires to perform invertion of a large matrix, also known as the KKR matrix, for each $k$ point in the irreducible Brillouin zone to obtain the $\tau^{nn}(\epsilon)$ matrix for a set of energy points, $\epsilon$'s, along a contour in the complex energy plane. The size of the KKR matrix is proportional to the number of atoms in the unit cell, and its inverse is essentially an $O(N^3)$ operation. As the number of atoms in the simulation cell (or unit cell) increases, a supercell approach based on KKR or other conventional \textit{ab initio} methods becomes less favorable, because of increasing computational cost. That is the large supercell sizes required for direct calculation of random disordered alloys make the standard \textit{ab initio} methods impractically expensive to use.  

While the mean field KKR-CPA approach described in the previous sections as well as the SQS method described below are designed to reduce the size of the simulation cell and are thus well suited for standard density functional method calculations, an alternative approach to the supercell method is to apply a real space scheme for solving the Kohn-Sham equation of density functional theory that will allow us to achieve linear scaling in the system size of the computational effort per self consistent iteration step as well as near optimal parallel performance on current high performance computing architectures. Specifically, we are using the Locally Self-Consistent Multiple Scattering (LSMS) method \cite{Wang1995}, which has demonstrated very good scaling and performance on contemporary computing architectures and has proved to be a very useful tool for performing the \textit{ab initio} calculation for complex structures involving tens of thousands of atoms~\cite{Eisenbach2017}. 

In the LSMS method, the linear scaling for the calculation of $\tau^{nn}(\epsilon)$ is achieved based upon an approximation that the multiple scattering processes involving atoms at a distance greater than a cut-off radius $R_{\rm LIZ}$ from atom $n$ are ignored. The idea behind this approximation is based on the observation that the scattering processes involving far away atoms influence the local electronic states less and less as the distance from the scatter under study is increased, an example of nearsightedness proposed by W. Kohn. In the LSMS method, the space within $R_{\rm LIZ}$ centered at an atom is called local interaction zone (LIZ) of the atom. If there are $M$ atoms in the LIZ centered at atom $n$, the computational cost for calculating $\tau^{nn}(\epsilon)$, and thus the GF, for atom $n$ does not depend on $N$, rather it depends on $M$. Since we only have to repeat the GF calculation for each atom, the total time cost for the entire electronic structure calculation will scale linearly with respect to $N$, the number of atoms in the unit cell.
The results obtained with this LSMS approach will converge with increasing the size of the LIZ. For sufficiently large LIZ sizes, the total energy result given by the LSMS method agrees very well with the KKR result, which is considered exact. Evidently, the LSMS method will allow us to readily perform calculation for supercells with thousands of atoms.

\subsubsection{The Special Quasi-Random Structure (SQS) Method
}
The Special Quasirandom Structure (SQS) is one of the realizations among all possible random structures of a periodic supercell. It is special in the sense that the particular structure provides the best approximation to the averaged local correlations \cite{Zunger_etal_1990,Wei_etal_1990}. The local correlations are defined in the context of the cluster expansion. A cluster is part of the lattice, the smallest cluster is one site of the lattice and the largest one is the entire supercell. Given a random realization of a binary alloy, the correlation of a cluster can be defined as the product of all sites within the cluster with sites occupied by element A designated as $+1$ and those by element B as $-1$. This correlation is averaged with respect to all the clusters obtained by point group symmetry and translation among all sites in a supercell.

Based on the assumption that the physical relevance of the correlation functions decrease with respect to the cluster size, in practice, only the correlations among the first few nearest neighbors are considered \cite{Wei_etal_1990}. The SQS is chosen as the realization of a supercell with correlation functions best matched the averaged correlation function for a set of chosen clusters. 
Instead of calculating all the random configurations for disorder realization averaging, the SQS method picks a configuration or a set of configurations which provides the best matched averaged correlation function as the representative for the random systems.


\section{Results}
To compare these methods, we have performed selfconsistent first principles calculations using our multiple scattering code MuST\cite {MuSTsite}. All the calculations employ the von Barth-Hedin approximation \cite{Barth1972} for the exchange correlation functional and muffin-tin potential type. For the KKR calculations, we use 60 special $k$ points. The LIZ size used for the calculations is 169. All calculations, both for the reciprocal as well as real space methods, use the same cutoff for $l$ of $l_{max} = 4$. The LSMS calculations for large random supercells include 1120 atoms in the supercell and the SQS calculations include 40 -- 160 atoms and are performed using the real space LSMS code using the same parameters as for the supercell calculations. We also ran simulations for different realizations of 40-atom SQS cells to assess the variations in total energy and density of states.  As can be seen in the insert in the right panel of fig.~\ref{fig:EvsCon}, The SQS total energy converges to the random supercell result and is already significantly closer to it than the mean-field CPA result for a calculation cell size of 40 atoms. The choice of the size of the local interaction zone (LIZ), the number of $k$ points and $l_{max}$ was determined from the convergence of the total energy of the system.
 As the CPA method does not readily account for the displacements of atoms from the ideal crystal site, we chose to also restrict the calculations for the supercell methods to ideal unrelaxed lattice sites, as the main aim of the present study is the comparison of these methods, and the origin of differences would have been obscured by the inclusion of displacements. It should be noted though,  that atomic displacements can play an important role in body centered high entropy alloys and these displacements can play an important role in stabilizing the bcc structure. \cite{Samolyuk2021}


\subsection{Ground State Properties: Total Energies, Formation Enthalpy, Density of States}

\begin{figure}
\includegraphics[width=0.5\linewidth]{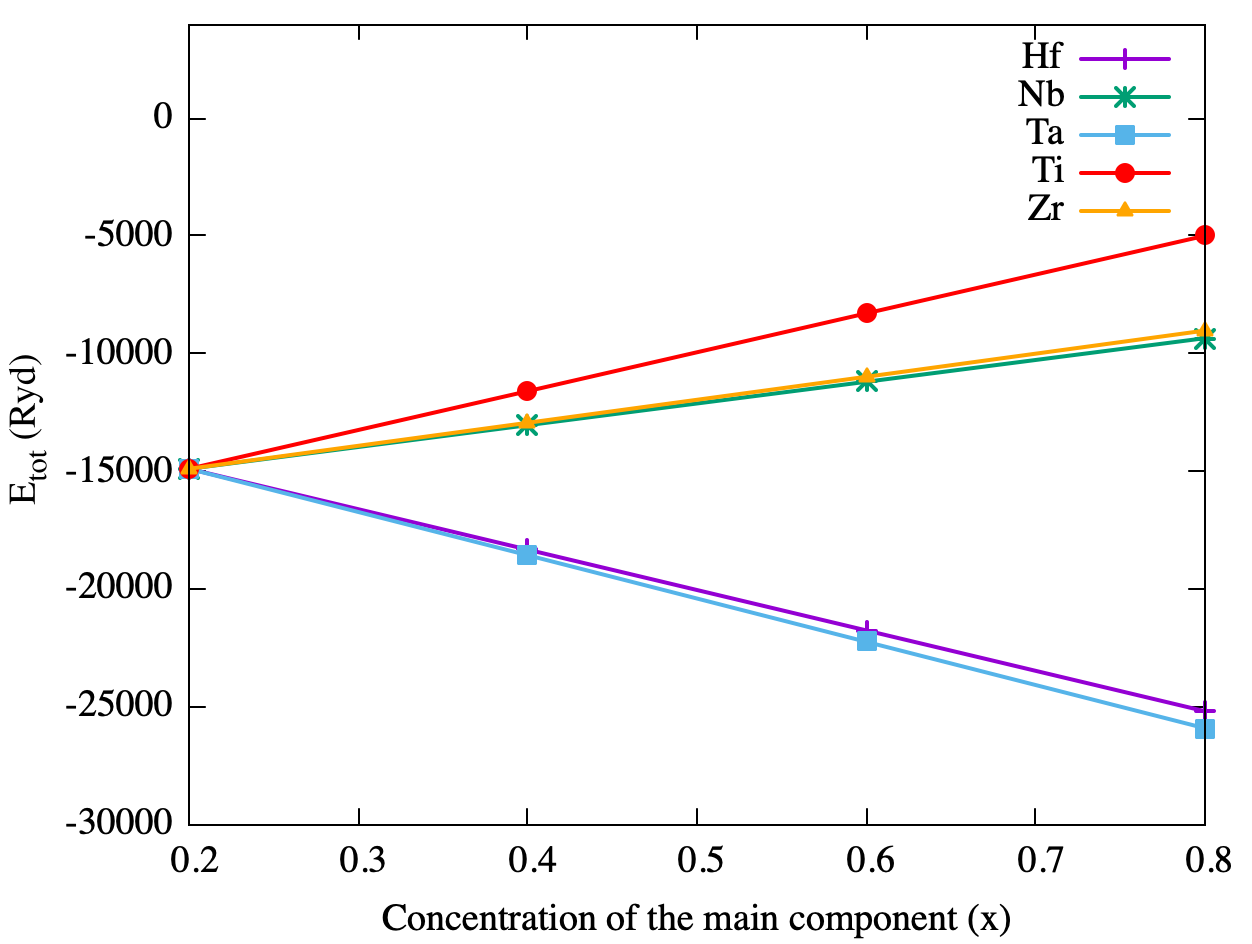}
\includegraphics[width=0.5\linewidth]{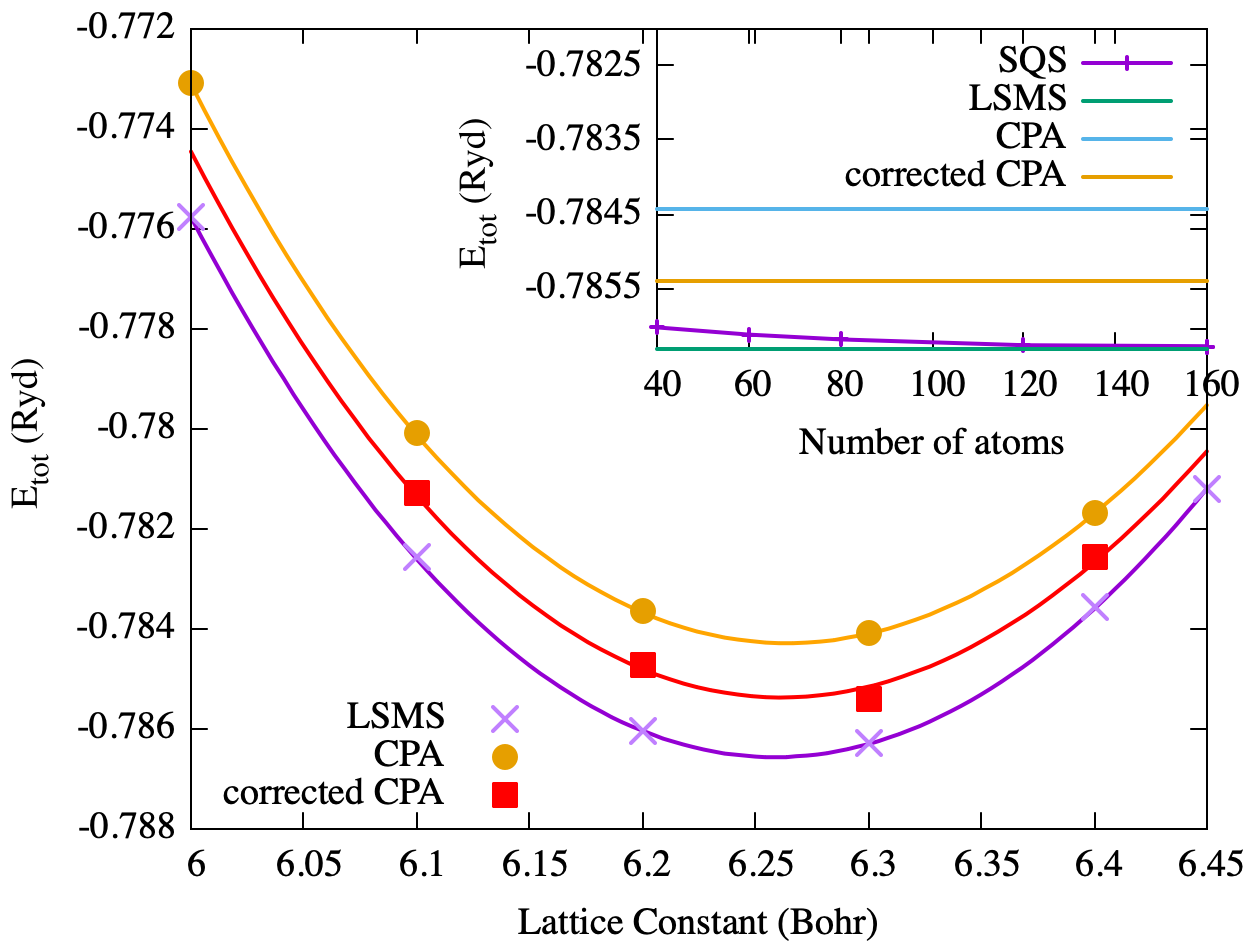}
 \caption{\label{fig:EvsCon} $\textbf{Left panel:}$ Total energy of the HfNbTaTiZr alloy from the LSMS method calculated for different concentrations of the components in the system. Labels correspond to the concentrations of the main component which was varied x, while the concentrations of the other components were kept as (1-x)/4.  LSMS system size is 1120 atoms.  Note that our calculations are all-electron methods, thus the total energy includes the binding energy of the core electrons, thus the grouping of the energy dependence on concentration into $3d$ (Ti), $4d$ (Nb, Zr) and $5d$ elements (Hf, Ta). $\textbf{Right panel:}$ Total energy - lattice constant dependence for the ${\rm Hf}_{0.05}{\rm Nb}_{0.05}{\rm Ta}_{0.8}{\rm Ti}_{0.05}{\rm Zr}_{0.05}$ system, with $E_{tot}$ offset of -25918 Ryd. Inset: Total energy of the ${\rm Hf}_{0.05}{\rm Nb}_{0.05}{\rm Ta}_{0.8}{\rm Ti}_{0.05}{\rm Zr}_{0.05}$ system from the LSMS, KKR-CPA, and corrected KKR-CPA methods as well as total energy for the different SQS system sizes. LSMS system size is 1120 atoms. $E_{tot}$ offset is -25918 Ryd. }
\end{figure}

The total energy was calculated for different concentrations of the components in the HfNbTaTiZr alloy for the same unit cell size, with 1120 atoms in total. The concentration of one component was varied $x$, while keeping the concentrations of other components $(1-x)/4$. This was done for all of the components in the HfNbTaTiZr alloy and the total energy was calculated for all of these variations (see Fig.~\ref{fig:EvsCon} (left panel)). 
The result shows that ${\rm Hf}_{0.05}{\rm Nb}_{0.05}{\rm Ta}_{0.8}{\rm Ti}_{0.05}{\rm Zr}_{0.05}$ alloy has the lowest total energy. Because of the restriction in the way to vary the concentration of each component here, there is a much larger configuration space that is not explored. Finding the global minimums that correspond to the stable states requires calculating the formation enthalpy of the alloys with unrestricted concentration variations. This is a computationally demanding task and it is beyond the focus of this paper.

The total energy of the Hf$_{0.05}$Nb$_{0.05}$Ta$_{0.8}$Ti$_{0.05}$Zr$_{0.05}$ system was calculated by all four methods: LSMS, SQS, KKR-CPA, and corrected KKR-CPA. The largest total energy difference is between LSMS and KKR-CPA methods, which is about 2.2 mRyd. For SQS, even with the system size of 40 atoms, the total energy difference between SQS and LSMS is less than 0.28 mRyd and keeps decreasing as the system size increases. Corrected KKR-CPA provides a lower total energy than the conventional KKR-CPA with the total energy difference between the corrected KKR-CPA and the LSMS being about 0.98 mRyd.

In Fig.~\ref{fig:EvsCon} (right panel), we show the calculated total energy, relative to an energy offset, of Hf$_{0.05}$Nb$_{0.05}$Ta$_{0.8}$Ti$_{0.05}$Zr$_{0.05}$ 
as a function of the lattice constant. As shown, the LSMS, KKR-CPA, and corrected KKR-CPA lattice constants agree with each other within the accuracy of the fit of the energy-lattice constant curve. The equilibrium lattice constant from all three methods for Hf$_{0.05}$Nb$_{0.05}$Ta$_{0.8}$Ti$_{0.05}$Zr$_{0.05}$ is about 6.26 Bohr and is lower than the experimental results for equimolar HfNbTaTiZr 6.42 Bohr \cite{se.sc.12,st.yu.18}, which is related to the smaller atomic size of Ta comparing to Hf or Zr \cite{zh.ji.19,mo.pe.19}.  As a matter of fact these curves can be collapsed, showing the excellent agreement between all these methods. For a given concentration we plotted on the right panel the decomposition of the energies on alloy components.  Given the similarities of the energies for a given concentration  we can expect that a similar collapsing is possible at other concentrations and the alloy components decomposition might be similar.
Note that the results correspond to the LSMS computation with more than 1000 atoms in the unit cell. Such computation is hardly possible in the SQS approach and the other methods, therefore we have taken the results from this computation only.

\begin{figure}[h]
\includegraphics[width=0.5\linewidth]{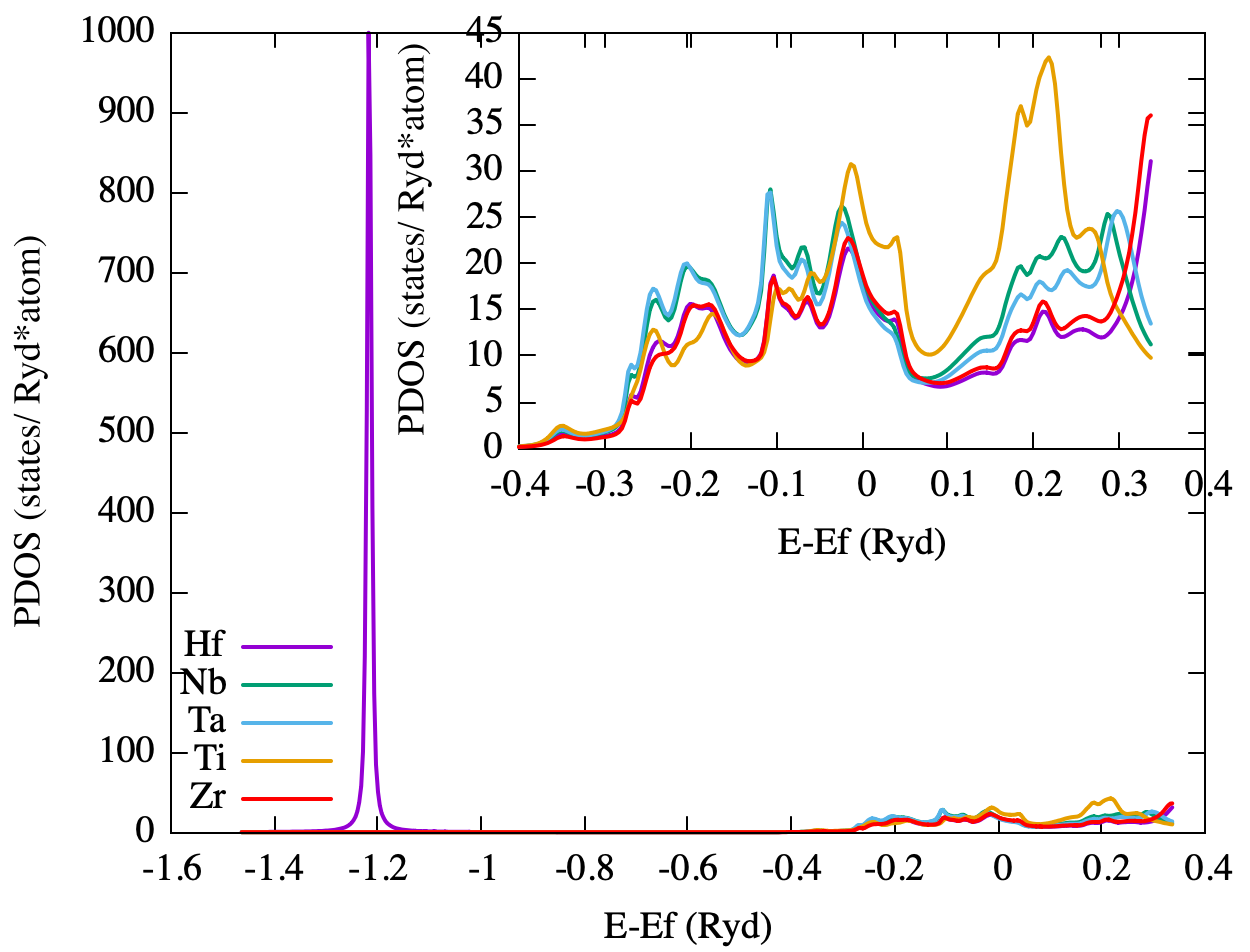}
\includegraphics[width=0.5\linewidth]{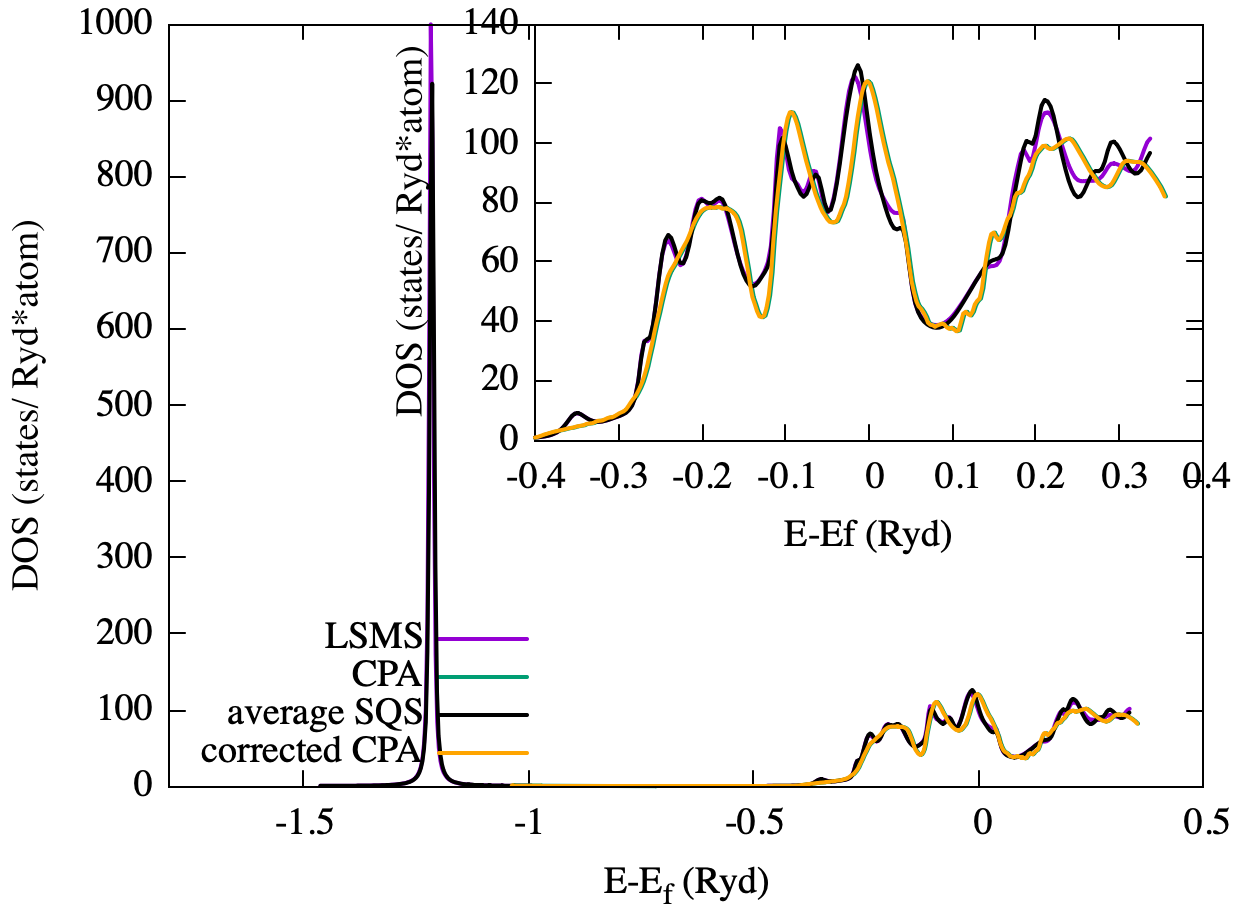}
\caption{\label{fig:DOS} $\textbf{Left panel:}$ Partial density of states from LSMS for all of the components of the Hf$_{0.05}$Nb$_{0.05}$Ta$_{0.8}$Ti$_{0.05}$Zr$_{0.05}$ alloy. Insert: the same figure, but for -0.4 to 0.4 Ryd E range.
 $\textbf{Right panel:}$ Total density of states for Hf$_{0.05}$Nb$_{0.05}$Ta$_{0.8}$Ti$_{0.05}$Zr$_{0.05}$ system from the LSMS, KKR-CPA, SQS and the corrected KKR-CPA methods. Insert: the same figure, but for -0.4 to 0.4 Ryd E range.  The
value of DOS at $E_F$ enters in the estimation of the electron-phonon
coupling in Sec. 3.3.}
\end{figure}

Next, we consider the partial and total density of states  normalized per atom, which we show in Fig.~\ref{fig:DOS}. The total density of states is in a good agreement for all four methods. In fact, there’s even better agreement between LSMS and SQS, and between KKR-CPA and corrected KKR-CPA (Fig.~\ref{fig:DOS}). The sharp peak on the left around -1.21 Ryd corresponds to Hf, as can be seen from the partial density of states in Fig.~\ref{fig:DOS} (left panel)~\cite{da.li.19}. The inserts are provided to better see the peaks between -0.4 and 0.4 Ryd.

As explained in section \ref{CPA-Correction}, the corrected CPA improves the total energy calculation by applying a screened charge model to include the long range Coulomb interaction effects. The screened charge correction essentially shifts the local potential for each species by an amount proportional to the local charge of the species. The major effect is on the total energy result, lowering the energy by $\sim 10$~mRyd if charge transfer is around 0.1$e$. Interestingly, the average DOS shown in Fig.~\ref{fig:DOS} is not sensitive to the screened charge correction. This can be understood by considering the fact that the average potential shift is zero, since there is only 1 atomic site per unit cell in the CPA approach and the average charge at each atomic site is 0.

To further compare our methods, we now perform the analysis of the formation enthalpy which is defined as:
\begin{eqnarray}
\label{eq:enthlpy}
\Delta E = && E_{{\rm Hf}_{(1-x)/4}{\rm Nb}_{(1-x)/4}{\rm Ta}_{x}{\rm Ti}_{(1-x)/4}{\rm Zr}_{(1-x)/4}} - ((1-x)/4)E_{\rm Hf} \\ \nonumber &&- ((1-x)/4)E_{\rm Nb} -  xE_{\rm Ta} - ((1-x)/4)E_{\rm Ti} - ((1-x)/4)E_{\rm Zr},
\end{eqnarray}
where $E_{\textrm{Hf}_{(1-x)/4}\textrm{Nb}_{(1-x)/4}\textrm{Ta}_{x}\textrm{Ti}_{(1-x)/4}\textrm{Zr}_{(1-x)/4}}$ is the total energy per atom of the alloy (table~\ref{tab:enthalpy}), and $E_{\textrm{Hf}}, E_{\textrm{Nb}}, E_{\textrm{Ta}}, E_{\textrm{Ti}}, E_{\textrm{Zr}}$ are the energies per atom of Hf, Nb, Ta, Ti, and Zr respectively. In table~\ref{tab:enthalpy} we show the formation enthalpies for the Hf$_{0.05}$Nb$_{0.05}$Ta$_{0.8}$Ti$_{0.05}$Zr$_{0.05}$ alloy calculated with LSMS, SQS, KKR-CPA, and corrected KKR-CPA methods. The formation enthalpies are negative, except for the KKR-CPA method (see table~\ref{tab:enthalpy}). This is not surprising since the conventional KKR-CPA method does not have the electrostatic potential energy properly included.  Negative formation enthalpy tells us that this alloy is energetically favorable comparing to the complete phase separation into the individual elemental constituents. Although, other phases are possible. Main arguments, such as negative formation enthalpy and high configurational entropy, suggest that this alloy is energetically favorable.

\begin{table*}[h]
\caption{\label{tab:enthalpy} Formation Enthalpies for Hf$_{0.05}$Nb$_{0.05}$Ta$_{0.8}$Ti$_{0.05}$Zr$_{0.05}$ in kJ/mol.  The SQS result is obtained from 40 atom supercell calculations.}
\begin{tabularx}
{\textwidth}{XXXX}
\hline \hline
 LSMS & SQS & CPA& corrected  CPA \\
\hline
-1.29214&-0.92588&1.60654&-0.10657\\
\hline \hline
\end{tabularx}
\end{table*}

Finally, we also calculated another experimentally relevant quantity, the bulk modulus , which was obtained using the Birch-Murnaghan equation of state by fitting the data from the KKR-CPA, LSMS, and corrected KKR-CPA. The results are in a fairly good agreement with each other and correspond to 158.4867 GPa from KKR-CPA, 158.1960 GPa from corrected KKR-CPA, and 159.0923 from the LSMS.The bulk modulus of Hf$_{0.05}$Nb$_{0.05}$Ta$_{0.8}$Ti$_{0.05}$Zr$_{0.05}$ is higher than the experimental results for equimolar HfNbTaTiZr, which is 134.6 GPa \cite{di.li.16}.





\subsection{Charge Transfer and Madelung Potential}

In alloys, the charge transfer takes place between atoms, driven by a combination of quantum mechanics and electrostatics. From the perspective of a computational approach based on DFT, the charge transfer occurs as a result of the charge self-consistency. For a periodic system with $N$ atoms per unit cell, the Madelung potential $V^n_{\rm mad}$, which is the electrostatic potential at atomic site $n$ due to the excess charges at all the other sites, is given by
\begin{equation}
\label{eq:Vmad}
    V^n_{\rm mad} = V^n_0 - e^2\sum^N_{m=1}M_{nm}\Delta Q^m,
\end{equation}
where $V^n_0$ is a constant potential, $M_{nm}$ is the Madelung constant\cite{PhysRevB.52.17106}, taking into account the fact that the charge distribution in the unit cell repeats itself in the entire space, and $\Delta Q^m$ is the excess charge (in the units of $e$) at site $m$. In the muffin-tin approximation (MTA), the electronic charge distribution in the unit cell consists of a spherical density distribution, $\rho^n(r_n)$, surround each atomic site within muffin-tin radius $R^n_{\rm mt}$ and a constant density $\rho_0$ in the interstitial region. The excess charge $\Delta Q^m$ in Eq.(\ref{eq:Vmad}) is defined to be the net charge in the atomic cell $\Omega_m$, 
\begin{equation}
    \Delta Q^m = Z^m - 4\pi\int_0^{R^m_{\rm mt}}dr_m\,\rho^m(r_m) - \rho_0\left(\Omega_m - \frac{4\pi}{3}\left(R^m_{\rm mt}\right)^3\right)
\end{equation}
where $Z^m$ is the positive charge of the nucleus, and the last term is the electric charge in the interstitial region inside the atomic volume. The constant potential in Eq.\ref{eq:Vmad} is given by
\begin{equation}
    V^n_0 = -e^2\rho_0\left[2\pi\left(R^n_{\rm mt}\right)^2+\sum_{m=1}^N M_{nm}\Omega_m\right].
\end{equation}
The terms in the square bracket in this expression depend purely on the geometrical parameters. If the atomic cell volume is chosen to be the same for all the atoms, which is a common practice in the KKR community and is also used through out the calculations carried out in this paper, by construction, the muffin-tin radius is the radius of the inscribed sphere radius of the atomic cell and is therefore also the same for all the atoms, and $V^n_0$ is essentially a constant, independent of atomic sites $n$. It is also necessary to point out that, in the atomic sphere approximation (ASA), there is no interstitial region, so that we have $\rho_0 = 0$ and $V^n_0 = 0$ in ASA~\cite{PhysRevB.66.024201}.

\begin{figure}[h]
\includegraphics[width=0.5\linewidth]{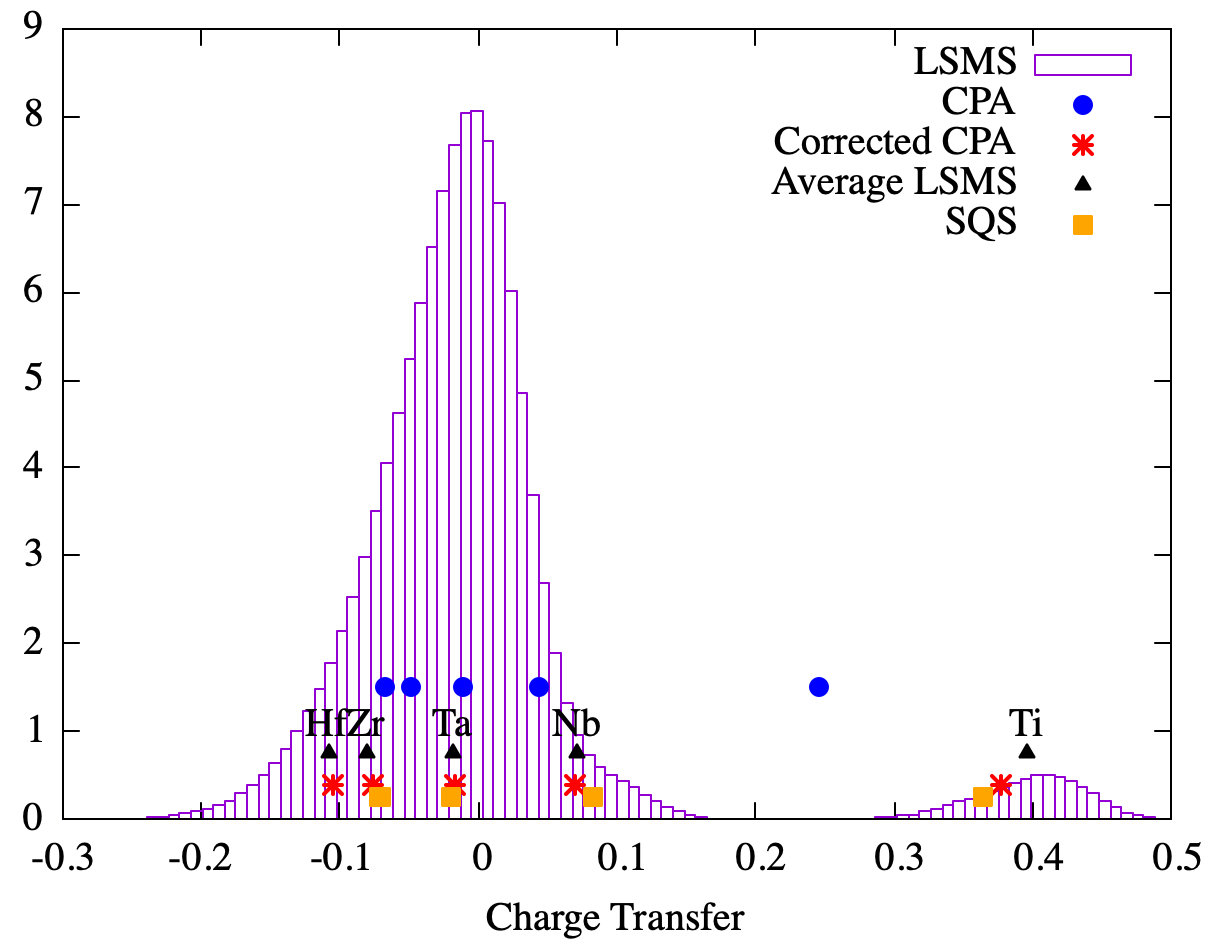}
\includegraphics[width=0.5\linewidth]{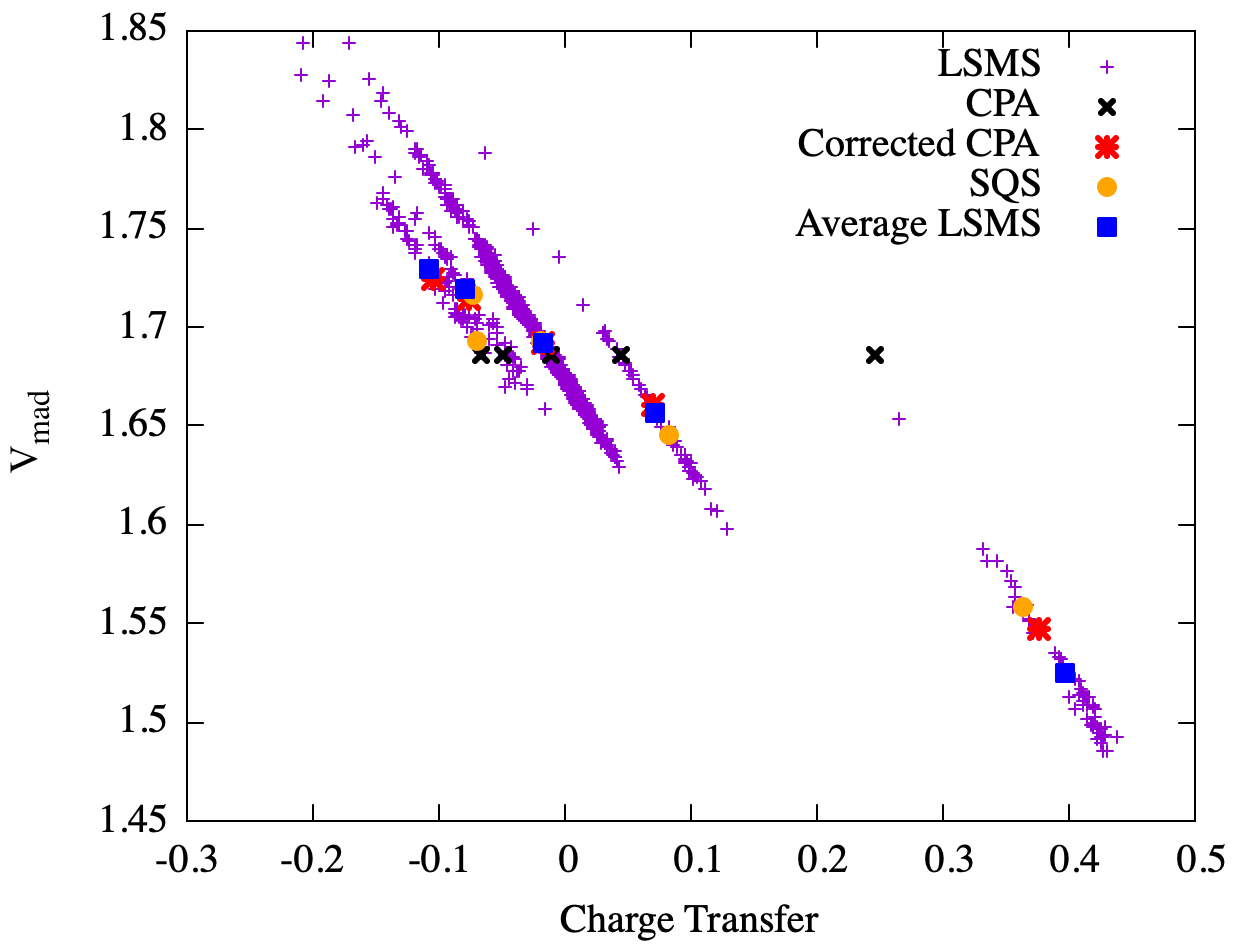}
\caption{\label{fig:chargeTr} $\textbf{Left panel:}$ Charge transfer distribution from LSMS method as well as the average charge transfer per atom species from LSMS, SQS, KKR-CPA, and corrected KKR-CPA methods for ${\rm Hf}_{0.05}{\rm Nb}_{0.05}{\rm Ta}_{0.8}{\rm Ti}_{0.05}{\rm Zr}_{0.05}$. $\textbf{Right panel:}$ Madelung potential $V_{\rm mad}$ versus charge transfer calculated by the LSMS, SQS, CPA, and corrected CPA methods for the ${\rm Hf}_{0.05}{\rm Nb}_{0.05}{\rm Ta}_{0.8}{\rm Ti}_{0.05}{\rm Zr}_{0.05}$ system.}
\end{figure}

\begin{figure}
 \includegraphics[width=0.5\linewidth]{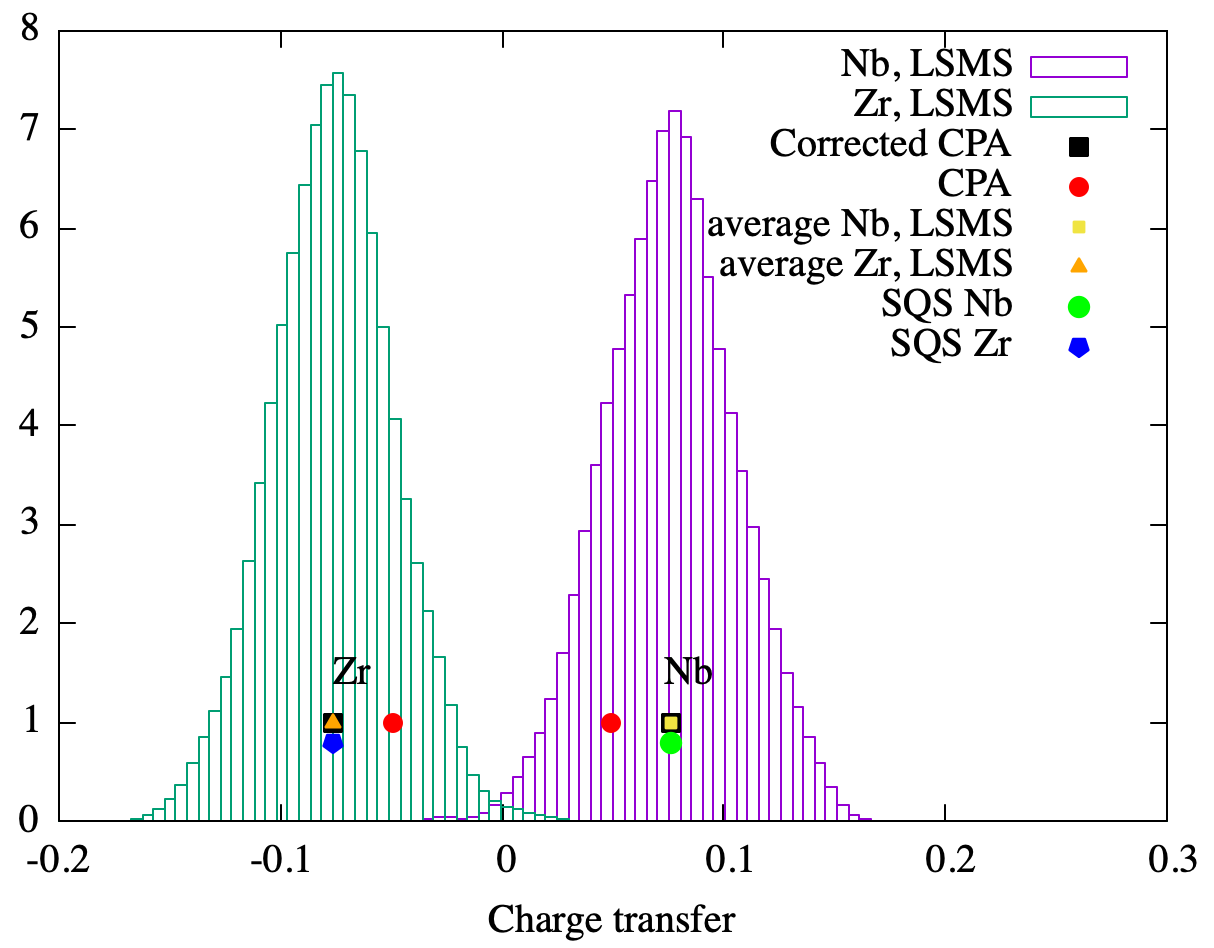}
\includegraphics[width=0.5\linewidth]{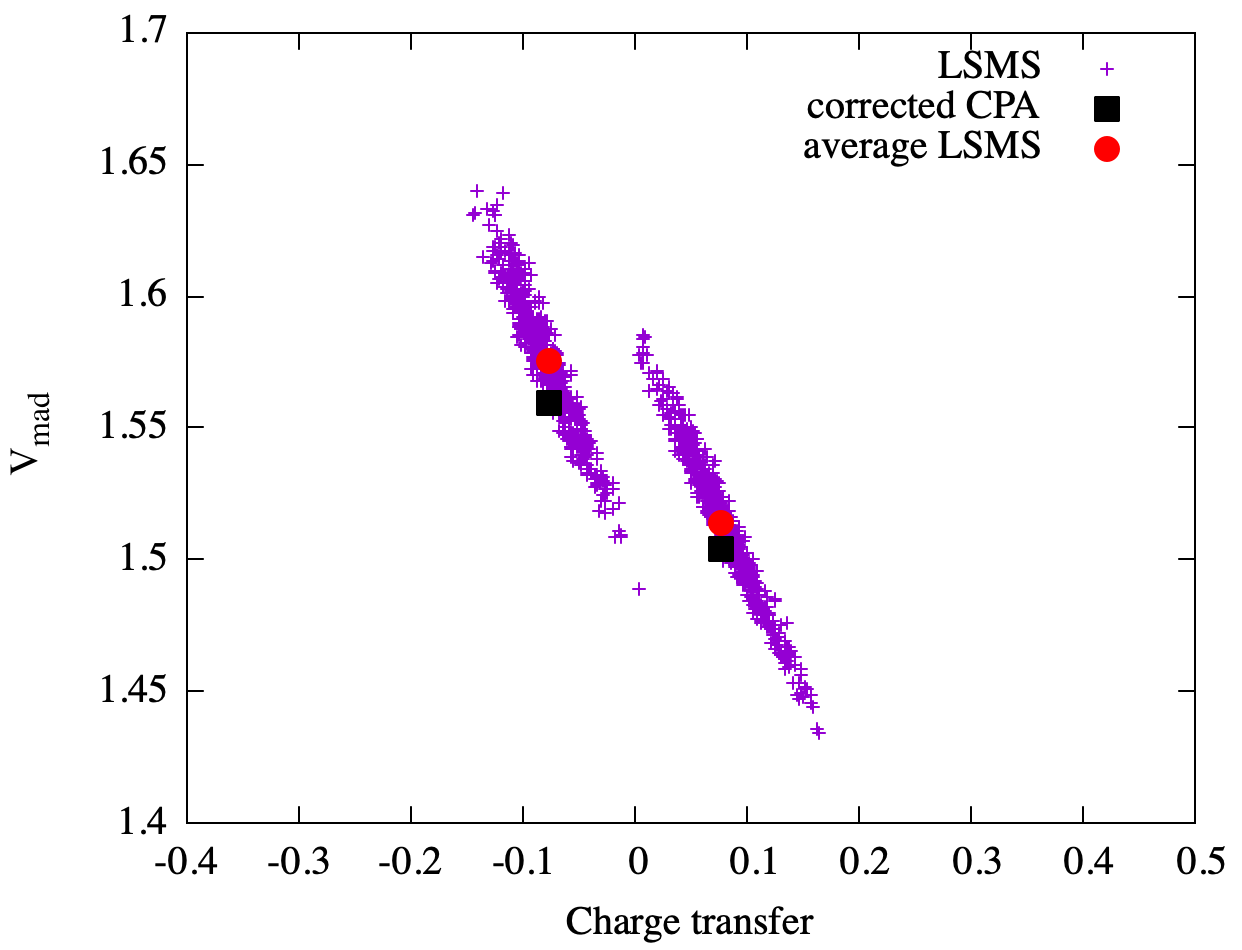}
 \caption{\label{fig:chargeTr_NbZr} $\textbf{Left panel:}$Charge transfer distribution per atom species from LSMS, KKR-CPA, corrected KKR-CPA, and SQS methods for equimolar ${\rm Nb}_{0.5}{\rm Zr}_{0.5}$. $\textbf{Right panel:}$ Madelung potential $V_{\rm mad}$ versus the charge transfer calculated by the LSMS and corrected CPA methods for the ${\rm Nb}_{0.5}{\rm Zr}_{0.5}$ system. }
 \end{figure}

For the study of random alloys, unlike local CPA method, the supercell calculation allows a direct observation of the charge transfer effects in the random alloy of a given configuration. A histogram of the charge distribution in HEA ${\rm Hf}_{0.05}{\rm Nb}_{0.05}{\rm Ta}_{0.8}{\rm Ti}_{0.05}{\rm Zr}_{0.05}$ and binary alloy NbZr is shown in Fig. \ref{fig:chargeTr} (left panel) and Fig. \ref{fig:chargeTr_NbZr}, respectively. The solid dots in the figures represent the averaged excess charge obtained by various methods. In the LSMS method in particular, the average is taken over all the atom of the same type in the entire unit cell which is usually chosen to be sufficiently large that self-averaging is assumed to takes place. Interestingly, despite the fact that the SQS method is also considered as a supercell approach, since its average is taken over a set of small unit cell samples, the average charge for Hf, Nb, Ti, and Zr in the ${\rm Hf}_{0.05}{\rm Nb}_{0.05}{\rm Ta}_{0.8}{\rm Ti}_{0.05}{\rm Zr}_{0.05}$ alloy shows small but noticeable difference from the averaged results given by the LSMS method, while its averaged charge for Ta agrees rather well with the LSMS. We speculate that as Hf, Nb, Ti, and Zr in this alloy have very low concentration, 0.05, the size of the samples for the SQS to take average over may not be sufficient. On the other hand, both LSMS and SQS agree very well on the average charge of Nb and Zr in the equimolar NbZr alloy case.   

It is necessary to point out that multi-element alloys also exhibit the qV relation, where the charge transfer at each site correlates almost linearly with the Madelung potential at the site, as shown by Fig. \ref{fig:chargeTr} (right panel), where the charge on the atoms of the same type falls on a straight line and the 5 visible lines formed by the crosses (LSMS results) in the figure are associated with 5 chemical elements. This linear phenomena, first observed by Faulkner et al.~\cite{PhysRevB.52.17106} for CuZn alloy, has been subjected to a number of investigations on various binary alloys~\cite{PhysRevB.55.7492,PhysRevB.57.15140,PhysRevLett.91.166401,PhysRevB.66.024201}. This relation is understood as a result of the charge screening effect in random alloys. The existence of the qV relation in HEAs allows to include the species dependent long-range electrostatic potential Eq. (\ref{eq:CPA_qV}) in the KKR-CPA calculations. The coefficients $A_\alpha$ and $B_\alpha$ in Eq. (\ref{eq:CPA_qV}) can be obtained from a supercell calculation.    

From the results shown in Figs. (\ref{fig:chargeTr}) and (\ref{fig:chargeTr_NbZr}), we also observe that the conventional CPA results for the averaged charge show a significant difference from the supercell (LSMS and SQS) results, especially for the charge on Ti in the HEA case. Not surprisingly, as explained in Sec.~\ref{CPA-Correction}, the average Madelung potential given by the CPA is the same for all elements. To overcome this deficiency of the CPA, we also performed the corrected CPA analysis by including a correction term, Eq. (\ref{eq:CPA_CS}), based on the charge screening model to the electrostatic potential for each chemical species in the charge self-consistency calculation. Both average charge and average Madelung potential results by the corrected CPA show a significant improvement over the conventional CPA results, and they agree well with the averaged LSMS results.



\subsection{Possible Superconductivity}

Superconductivity in the Hf$_{0.08}$Nb$_{0.33}$Ta$_{0.34}$Ti$_{0.11}$Zr$_{0.14}$ alloy has been reported a few years ago~\cite{ko.vr.14}. The alloy has a bcc structure and the reported critical temperature of $T_c=7.3$~K is slightly smaller than that of the pure Nb metal ($T_c = 9.2$~K) and to the NbTi
alloys ($T_c \approx 10$~K), which are the mostly-used practical superconductors. In the reported superconducting HEA specific heat measurement confirmed bulk superconductivity and the conventional phonon mediate pairing was inferred. 
An estimate of the critical temperature in the conventional superconductors can be obtained by knowing the magnitude of the electron-phonon couplings.
However, electron-phonon calculations in disordered alloys from first principles remain computationally expensive, therefore phenomenological approaches like the  Gaspari-Gyorffy theory~\cite{ga.gy.72} still provide the input for these estimates. 

In the following we briefly go through the essential ideas of the 
Gaspari-Gyorffy approach~\cite{ga.gy.72}. 
Within the theory of strongly-coupled superconductors~\cite{mcmi.68} the electron-phonon coupling constant $\lambda$ can be
expressed as: 
$$\lambda=N(E_F)\langle I^2 \rangle/M \langle \omega^2 \rangle , $$
where $N(E_F)$ is the DOS at the Fermi level, $\langle I^2 \rangle$ is the electron-phonon matrix element, averaged over the Fermi surface, $M$ is the
atomic mass, and $\langle \omega^2 \rangle$ is the average squared phonon frequency. The numerator can be rewritten as
$\eta=N(E_F)\langle I^2 \rangle$, which is known as the Hopfield parameter.
In a so-called local-phonon representation the electron-phonon interactions mainly consist of scatterings that
change the electronic angular momentum $l$~\cite{hopf.69}. 
Using these ingredients Gaspari and Gyorffy
proposed an approximate way to compute 
$\langle I^2 \rangle$ using the multiple-scattering Green’s function formalism and adopting the
rigid muffin-tin approximation.
In these approximations the Hopfield parameter is computed from a combination of electronic scattering phase shifts and
the electronic densities of states. 
The average squared phonon frequency $\langle \omega^2 \rangle$ can in principle be computed from the phonon density of states, but since this is hard to obtain for disordered systems, it can be approximated by using the Debye temperature $\theta_D$ via $\langle \omega^2 \rangle \approx \frac{1}{2} \theta^2_D$.
Bringing everything together, the electron-phonon coupling constant can be computed as $\lambda=\langle \eta \rangle/\frac{1}{2}\langle M \rangle\theta^2_D$, where $\langle \eta \rangle$ and $\langle M \rangle$
are the disorder-averaged Hopfield parameter and atomic mass, respectively. 

In the present work, we computed the Hopfield parameter within the exact muffin-tin orbitals (EMTO) method~\cite{an.je.94.2,vi.sk.00,vito.01}. The atomic masses were taken from Ref.~\cite{Jas2016}. Since we do not know the Debye temperature for Hf$_{0.05}$Nb$_{0.05}$Ta$_{0.8}$Ti$_{0.05}$Zr$_{0.05}$, we used the experimentally determined Debye temperature for Hf$_{0.08}$Nb$_{0.33}$Ta$_{0.34}$Ti$_{0.11}$Zr$_{0.14}$~\cite{ko.vr.14}, $\theta_D=243$~K. 
This gives us an electron-phonon coupling $\lambda=0.9$. 
%
McMillan~\cite{mcmi.68} provided a solution for the finite temperature Eliashberg theory and found the $T_c$ for various cases as an approximate equation relating $T_c$ to a small number of simple parameters:
\begin{equation}
    T_c = \frac{\theta_D}{1.45} \mathrm{exp} \left[ -\frac{1.04(1+\lambda)}{\lambda-\mu^{*}(1+0.62\lambda)} \right],
\end{equation}
The $1+\lambda$ plays the role of electron mass enhancement and the parameter $\mu^{*}=0.13$ is an effective Coulomb repulsion reflecting the retardation effect of electron-phonon coupling with respect to the instantaneous Coulomb repulsion~\cite{mcmi.68}. With our computed $\lambda$, we get a critical temperature $T_c=9$~K. It should be noted that the Gaspari-Gyorffy theory overestimated $T_c$ for Hf$_{0.08}$Nb$_{0.33}$Ta$_{0.34}$Ti$_{0.11}$Zr$_{0.14}$ by roughly a factor 2, therefore it seems likely that a similar overestimation would be the case also for the presently investigated alloy.

 In section 3.1 the results of DOS computations are presented. Fig. 2 on the left the alloy component DOS is seen and on the right panel the total DOS in the different modeling of disorder is presented. In what superconductivity is concerned, we used the McMilan formula ($\eta$ and $\lambda$ parameters) in which the DOS contributes through its value at the Fermi energy. From the total DOS, we see that around $E_F$ all methods (LSMS, SQS, CPA, and corrected CPA) produce  similar values for the total DOS at $E_F$. Thus, the estimated $T_c$ in all approaches will be similar.

\section{Conclusion}\label{sec13}

In this paper we provided a comparison of different first principles approaches for treating disordered substitutional alloy systems. While the main focus was on multicomponent bcc refractory metal alloys, we expect that our general conclusions are applicable to a wider range of alloys. The methods that we have compared fall into two categories, namely the mean field descriptions of chemical disorder, as represented by CPA and corrected CPA, and supercell representations of disorder, as represented by the SQS approach and large random supercell LSMS calculations. As a large random supercell will reduce the spurious periodic interactions of sites with themself and also capture the full distribution of atomic neighborhoods as the cell size goes to infinity, this approach represents the standard against which the other approaches should be evaluated.
In this comparison, the straight CPA method clearly falls short in capturing some physical properties, including the wrong sign for the enthalpy of formation. We have identified the source of this discrepancy as originating from the assumption of a constant Madelung potentials for all atomic species. The atomic species specific screening correction for this deficiency significantly improves the mean field results and allow us to draw conclusions of the qualitative behavior. The SQS method has very good agreement with the random supercell method and approaches the results with increasing SQS cell sizes.

Thus, in many practical cases the overarching factor to decide which method to employ will depend on the computational resources available as well as the number of different concentrations to be investigated.
The CPA approach has the smallest resource requirements as it can represent the alloy on the minimal unit cell of the underlying crystal lattice. This will allow high volume screening of many compounds and concentration and it also allows for arbitrarily small fractional site concentrations that are not readily representable in supercell methods without resorting to very large cell sizes. For higher accuracy requirements, the SQS method is likely the most practical approach for traditional DFT methods that scale with the cube of the system size, yet the construction of the SQS atom arrangements for large cells and averaging over potential ensembles of cells can also represent a non-negligible cost.
For linear scaling methods, such as our LSMS approach, the random supercell approach has the advantage of minimizing the assumptions that need to go into constructing the simulation cells and it allows for the study of the distribution of local quantities, such as the charge distribution, that cannot be captured by smaller systems. It also has the added potential to investigate arbitrary atomic arrangements while minimizing spurious periodicity induced artifacts.

\backmatter

\section*{Acknowledgements}
We thank Dr. G. Malcolm Stocks for valuable discussions of first principles alloy theory.
The work of ME (LSMS code development) was supported by the U.S. Department of Energy, Office of Science, Basic Energy Sciences, Materials Science and Engineering Division. 
LC and A\"O acknowledges the financial support by the Deutsche Forschungsgemeinschaft
through TRR80 (project F6) Project number 107745057.
This research used resources of the Oak Ridge Leadership Computing Facility, which is supported by the Office of Science of the U.S. Department of Energy under Contract No. DE-AC05-00OR22725.
This work is based on open-source ab initio software package MuST~\cite{MuSTsite}, a project supported in part by NSF Office of Advanced Cyberinfrastructure and the Division of Materials Research within the NSF Directorate of Mathematical and Physical Sciences: HT and WM acknowledge NSF OAC-1931367; KMT acknowledges NSF OAC-1931445; and YW acknowledges NSF OAC-1931525. Work in Florida (WDH and VD) was partially supported by the NAF grant No. DMR-1822258.

\section*{Declarations}



\bmhead{Conflict of interest} The authors declare that they have no conflict of interest.

\bmhead{Availability of data} The data from our calculations is available on request.

\bmhead{Code availability} The MuST suite of codes is available at \href{https://github.com/mstsuite}.

\newpage

\bibliography{master}


\end{document}